# Title: A novel approach to profile global circulation pathway of SARS-CoV-2 variants by site-based mutation dynamics

**Authors**: Hong Zheng[1,2], Shimin Su[1,2], Caiqi Liu[1,2], Jingzhi Lou[1,2], Lirong Cao[1,2], Yexian Zhang[1,3], Zhihui Zhang[1,2], Marc Ka Chun Chong[1,2], Benny Chung-Ying Zee[1,2], Peter Pak-Hang Cheung[4], Haogao Gu[5], Juan Pu[6], Leo Lit Man Poon[5,7,8], Hui-Ling Yen[5], Maggie Haitian Wang[1,2*]

[1] Center for Clinical Research and Biostatistics, JC School of Public Health and Primary Care (JCSPHPC), the Chinese University of Hong Kong (CUHK); Hong Kong SAR, China.

[2] CUHK Shenzhen Research Institute, Shenzhen, China.

[3] Beth Bioinformatics Co. Ltd; Hong Kong SAR, China.

[4] Department of Chemical Pathology, CUHK; Hong Kong SAR, China.

[5] School of Public Health, Li Ka Shing Faculty of Medicine, The University of Hong Kong; Hong Kong SAR, China.

[6] National Key Laboratory of Veterinary Public Health and Safety, Key Laboratory for Prevention and Control of Avian Influenza and Other Major Poultry Diseases, Ministry of Agriculture and Rural Affairs, College of Veterinary Medicine, China Agricultural University, Beijing 100193, China

[7] HKJC Global Health Institute, LKS Faculty of Medicine, The University of Hong Kong, Hong Kong SAR, China.

[8] HKU-Pasteur Research Pole, School of Public Health, LKS Faculty of Medicine, The University of Hong Kong, Hong Kong SAR, China.

[*] Correspondence: maggiew@cuhk.edu.hk (M.H.W)

## Abstract

The genetic evolution of SARS-CoV-2 has caused recurring epidemic waves, understanding its global dispersal patterns is critical for effective surveillance. We developed the Site-based mutation dynamics – Equal Power Sampling (S-EPS) framework, a phylogenetic-free, bias-correcting framework for profiling viral source-sink dynamics. Applying S-EPS to 6.6 million SARS-CoV-2 genomes (March 2020 - June 2024) from 13 regions worldwide, we identified Africa and the Indian subcontinent as the predominant sources of key mutations. Southeast Asia serves as an early transmission hub, while Russia and South America mainly acted as sinks. Key mutations took longer to establish fitness in source regions than externally. Once an amino acid substitution on the receptor-binding domain reached 1% prevalence in major sources, there is an 80% probability it would spread elsewhere, with a 2-month median lead time (IQR: 1-4). Our findings underscore the importance of genetic surveillance, with S-EPS offering enhanced capability for monitoring emerging viral threats.

The SARS-CoV-2 virus continuously evolves, generating genetic variants that drive recurrent waves of COVID-19 epidemics[1]. Profiling the global circulation pathway of the virus is essential for understanding its genetic epidemiology and enabling early warning of emerging variant threats. Previous studies have largely focused on identifying geographical regions where variant exportation exceeds importation[2], or on characterizing the dissemination patterns of specific variants within continents or countries following initial introduction[3-7]. Currently, the presumed origin of a variant is still regarded as the country where the first outbreak associated with that variant was reported. Determining the actual source regions is challenging due to uneven data sampling, as countries with more intensive surveillance efforts generate larger volumes of genomic data, which can disproportionately influence phylogenetic reconstruction and obscure the role of other regions[8]. In this study, we characterize the source-sink dynamics of SARS-CoV-2 from a novel perspective by analyzing site-based mutation dynamics using an Equal Power Sampling (S-EPS) framework. This approach mitigates sampling bias by estimating mutation prevalence with equal statistical power in two stages across all geographical regions. Therefore, the framework infers the relative position of each region along the global transmission timeline and traces the geographical sources contributing to the formation of genetic variants.

To elucidate global source-sink dynamics of SARS-CoV-2 variants, we analyzed 6.6 million viral genomes collected between March 2020 and June 2024, representing 44.6% of all sequences in the GISAID database[9] for that period. The analysis spans 13 regions across six continents, chosen for their longitudinal data completeness and utility as continental proxies, covering Asia (Israel, India, Singapore, Japan), Africa (South Africa), North America (California, New York, Mexico), South America (Brazil), Europe (United Kingdom, Germany, Russia), and Oceania (Australia). By integrating geographical regions and monthly temporal solution, we defined 676 space-time data strata to model the global dynamics of viral mutations (**Table S1**). We then calculated the prevalence of 170 key substitution mutations characterizing the SARS-CoV-2 variants[10], including Alpha, Beta, Gamma, Delta, and Omicron sub-lineages (BA.1/2, BA.4/5, BA.2.75, XBB, and JN.1). Each mutation was both geographically widespread (present in >50% of the regions) and dominant in at least one region (**Table S2**, **Methods**).

**The most probable source regions of SARS-CoV-2 key mutations and the process of variant formation**

Using the S-EPS framework (**Methods**), we first calculate rank statistics for geographical regions based on the timing of each key mutation's emergence to infer its transmission pathway. For instance, the location where a key mutation was first detected is ranked No.1, and the last location is ranked No.13. By treating each mutation as an "experiment" and aggregating data across all key mutations and repeated random samples, we can derive the probability distribution of regions' ranks throughout the virus's circulation pathway.

Consequently, the most probable source region is identified as the one with the highest probability of being ranked No. 1 for the initial observation of key mutations. This analysis was performed for individual variants and then aggregated across all variants (**Fig 1, Supplementary Table S3 and S4**). For the Alpha variant, the most probable source is the United Kingdom with a probability of 52.9% (**Fig 1a**), for the Delta and Omicron XBB, India (68.4%, 46.7%) (**Fig 1b** and **1d**), and for the Omicron BA.1/2 and JN.1, South Africa (59.5%, 48.8%) (**Fig 1c** and **1e**). Across all variants, the top two most probable genetic sources are South Africa (32.0%) and India (25.8%), which together contribute approximately 60% of all key SARS-CoV-2 mutations (**Fig 1f**). While our results coincide with the presumed origin of individual variants, to our knowledge, this is the first model that statistically identifies the geographical source of major SARS-CoV-2 variants. Notably, 30-50% of the key mutations for any given variant originated from outside the initial outbreak region. Across all variants, the United States contributed 8.7% (California: 5.5%, New York: 3.2%), United Kingdom 7.4% and Brazil 5.8% (**Fig 1f, Table S3**). These results suggest that the advantageous, variant-defining amino acid changes in SARS-CoV-2 are not formed in a single region. Although convergent evolution may have occurred, the possibility of concurrent transmission and evolution cannot be excluded, as a substantial proportion of positively selected mutations accumulated during global circulation. Nevertheless, the primary source region played a critical role by contributing the largest and final batch of advantageous mutations, leading to the formation of a major variant and a subsequent local epidemic outbreak.

This process of forming variants is illustrated in **Fig 2**. Using Omicron BA.1/2 as an example (**Fig 2a**), the variant was designated as a Variant of Concern (VOC) by the World Health Organization (WHO) in February 2022. Of its 28 key mutations in the S protein, 25% (7/28) were first detected in regions outside the initial outbreak location, South Africa. These includes S477N, initially identified in Australia in March 2020; T95I, first detected in New York in July 2020; and N679K, first observed in Brazil in August 2020. As time approached June 2021, the remaining amino acid changes in the S protein of Omicron BA.1/2 emerged in South Africa, with 15 of these appearing in October 2021, just before the Omicron outbreak in November 2021. A similar pattern is observed across all major SARS-CoV-2 variants including Alpha, Beta, Delta, Omicron BA.1/2, XBB and JN.1 (**Fig 2b**). In addition, we note that the diversity of geographical regions contributing key mutations has increased in more recent variants such as XBB and JN.1.

**Global circulation pathways of the SARS-CoV-2**

Using rank probabilities, we further identified major intersection regions during the early spread of virus. We define an early intersection as a region where key mutations were detected shortly after their emergence in the source location, specifically, when they appeared in rank positions 2 or 3 along the global circulation pathway. Our analysis shows that the most probable early intersections are generally geographically proximate to the most likely source of each genetic variant (**Fig 1**). For instance, for the Alpha variant, the most probable early intersection was Israel (60.8%), following its emergence in the UK (**Fig 3**). For the Delta variant, early intersections included Australia (34.1%) and Singapore (27.7%), after the variant was first detected in India. For Omicron BA.1/2 variant, India (17.2%) was the primary early intersection following its origin in South Africa. For XBB, Singapore (34.1%) served as the early intersection after India. For JN.1 variant, the UK (13.8%) and Germany (13.1%) were early intersections following South Africa. Across all variants, the most frequent early intersection regions after a variant left its source were Singapore (13.9%), Australia (10.7%), India (10.6%), the UK (10.0%), and Israel (9.5%) (**Fig 3**). In contrast, the most probable sink regions, where key mutations were identified at the latest stages of transmission, were Russia (17.5%) and Brazil (16.2%).

**Prediction of future predominant mutations leveraging the source region data**

We investigated whether data from a virus's source region could forecast future predominant mutations. To evaluate this, we considered amino acid changes in the receptor binding domain (RBD) that reached a prevalence of 1% in the top source regions (South Africa and India) as a prediction set. The conditional probability that these source region-derived mutations would become predominant in at least one of the other 11 geographical regions was 80.0%, with a median prediction lead time of 2 months (interquartile range – IQR: 1–4 months). When using a 2% prevalence threshold for amino acid changes across the entire spike protein in the source regions, the probability of these mutations becoming predominant elsewhere was 73.6%, with a median lead time of 2 months (IQR: 1–3 months). Results under alternative thresholds are provided in **Supplementary Table S5**. These findings underscore the critical importance of persistent global genome surveillance for the timely detection and prediction of emerging variants.

## Discussion

This study presents the novel S-EPS framework for delineating the source-sink dynamics of SARS-CoV-2 variants. The method is robust to substantial spatiotemporal variations in genomic sample size, making it suitable for tracking fast-evolving and rapidly spreading pathogens. Through the analysis of extensive genome data, we identified key sources, intersections, and sinks of these variants. Moreover, we showed that the source region data can forecast future SARS-CoV-2 mutations with a high likelihood of emergence and eventual dominance.

**Evolutionary traits of SARS-CoV-2 in the estimated source regions**

To better understand the evolutionary dynamics of SARS-CoV-2 virus, we analyzed the mutation transition time—defined as the period required for a key mutation to emerge and become predominant in a specific region (**Methods**)[11]. Our results indicate that the median transition time for key mutations in the estimated source region was 7 months (IQR: 5–11 months), which is significantly longer than the median of 5 months (IQR: 3.5−6 months) observed in the non-source regions (Wilcoxon $p < 0.001$; **Fig 3c**). Additionally, genetic diversity was higher in the source region, as reflected by the number of mutations reaching 1% prevalence: the median in the source was nearly 50% higher than in the non-source regions (Wilcoxon $p = 0.051$). These findings suggest that the greater genetic diversity in the source region might have facilitated the emergence of novel variants with selective advantages, a process likely requiring extended time frame. Once a variant with full fitness arises, it can spread rapidly to other regions and sweep through populations in a relatively short period.

### Shared ecological traits of the source regions for SARS-CoV-2 and influenza A/H3N2

We observed that the most probable source regions for SARS-CoV-2 variants, South Africa and India, share similar ecological traits, specifically, high population density and tropical or subtropical climates, whereas the major sink region, Russia, lies predominantly within the temperate zone. This ecological characteristics mirrors that of seasonal influenza virus A/H3N2, in which new lineages typically originate in tropical or subtropical regions of East and Southeast Asia before migrating to temperate regions[12,13], which we previously identified using a preliminary site-based model[14]. Interestingly, SARS-CoV-2 and H3N2 also share similarities in their global dissemination, rapid evolutionary rates, and strong selective pressures. The combination of high population density and warm climates likely facilitates sustained viral circulation[14,15], resulting in less stringent population bottleneck and increased genetic diversity. These conditions may in turn promote the emergence of lineages with selective advantage.

### Advantages of the site-based framework, what it estimates and what it does not estimate

The site-based framework proposed in this study offers several advantages for analyzing rapidly evolving, highly transmissible, and genetically diverse pathogens. First, it enables the identification of complex genetic source regions when a variant's characteristic mutations are inherited from multiple populations during its spread. Second, by estimating mutation prevalence and applying an equal-power sampling strategy, the method mitigates biases arising from uneven surveillance capacities across countries and over time, thereby allowing equitable comparison of emerging mutations across time and space. A third advantage lies in the method's robustness to variations in viral transmission speed, achieved through the use of rank statistics. Although factors such as transmissibility and public health interventions can affect the absolute timing and pace of viral spread[16], the rank statistic reliably captures the sequential order in which key mutations are observed worldwide. Fourth, the framework is computationally efficient and scalable to large genomic datasets. For instance, processing one million SARS-CoV-2 genomes required only 6.2 hours on a standard laptop. This computational performance makes it feasible to analyze the largest genomic datasets available to date, offering comprehensive coverage of global SARS-CoV-2 data and supporting real-time surveillance and updates on emerging variants. It is important to note that the primary objective of this site-based framework is to estimate the rank statistics of geographical regions along the viral migration pathway. Unlike phylogenetic models, which aim to reconstruct strain genealogies and can be sensitive to sampling disparities, this framework does not infer strain ancestry or the precise timing of a new variant's introduction. Instead, it focuses on elucidating the relative positions of regions throughout the viral spread process.

**Genetic origins versus dissemination hubs**

A key distinction of this study is its focus on identifying the genetic source regions of SARS-CoV-2, in contrast to prior research that primarily located the virus's major dissemination hubs. For instance, Tegally et al. highlighted countries with the highest net exportation and importation of variants, identifying them as key hubs for the global spread of SARS-CoV-2 variants[2]. Their findings suggested that the virus's presumed origin in global dissemination diminished over time compared to that of travel hubs. They reported the United States as the largest contributor to viral exports, followed by India, the United Kingdom, South Africa, and Germany[2]. In contrast, our study identifies the *genetic source regions* that contributed to the emergence of new variants. Our analysis further reveals that the early intersection regions of viral spread are geographically closer to these genetic regions than to the other areas. This observed migration pattern reflects the sequential global emergence of variants.

**Limitations of the study**

This study analyzed 13 geographical regions across six continents that maintained continuous and sufficient SARS-CoV-2 sequencing data. A general limitation of any computational approach is that new mutations may emerge from unsequenced areas. Nevertheless, as our results demonstrate, the regions identified as the most probable sources are those where the final set of key mutations arose, indicating that our analysis successfully captured the major genetic sources of new SARS-CoV-2 variants. Expanding systematic genetic surveillance to more geographical regions over time will further refine the profiling of global virus circulation pathway. Some regions included in the analysis had limited sample sizes, which may have led to an underestimation of their priority in the global circulation pathway. However, South Africa, which had the most insufficient sampling, was still identified as one of the most probable sources, mitigating this concern to some extent. A further limitation is that our study focused exclusively on amino acids substitutions; we did not account for other types of mutations, such as synonymous nucleotide changes, insertions, or deletions. are not accounted. Finally, the source-sink patterns we identified are based on the historical behavior of the SARS-CoV-2 virus. Whether future variants will adhere to similar circulation patterns requires continued monitoring.

In summary, this study introduces a novel analytical framework to investigate the source-sink dynamics of rapidly evolving pathogens and to trace the genetic origins and migration routes of SARS-CoV-2 variants. This framework can be readily deployed to detect emerging advantageous mutations, providing critical insights for vaccine design and epidemic preparedness. By identifying source regions and maintaining systematic genome surveillance, our ability to predict future vaccine strain can be markedly enhanced.

## Methods

### Key mutations of SARS-CoV-2 variants

We begin by defining what a mutation is, followed by the criteria for identifying key mutations. A mutation is identified when an amino acid residue at a specific position in the genome differs from the corresponding residue in the original SARS-CoV-2 sequence, hCoV-19/Wuhan/Hu-1/2019 (GISAID accession: EPI_ISL_402125). Let $X_k$ represent a mutation in amino acids (AA), where $1 \leq k \leq K$ and $K = J \times 19$. Here, $J$ denotes the length of the AA sequence under consideration, and 19 represents the number of possible AA outcomes. Let $p_{kl}(t)$ denote the prevalence of mutation $X_k$ in region $l$ at time $t$, where $l$ =1, ..., $L$ and $t$ = 1,..., $T$. In this study, $L$ = 13 represents the total number of geographical regions analyzed, and $T$ = 52 is the number of time intervals in the investigation period. A mutation is classified as a *key mutation* if it satisfies the following conditions: (1) **Characteristic mutation**: The mutation has been designated as a characteristic mutation of SARS-CoV-2 variant, as defined in WHO reports[10]. (2) **Fitness**: The mutation must be predominant in at least one region, meaning there exists at least one region $l$ where $p_{kl}(t) \geq 0.5, l \in L$. (3) **Coverage**: The mutation must be observed with a prevalence exceeding 20% in more than half of the $L$ regions (i.e., at least 7 regions in this study). Let $L_k$ represent the number of regions where $p_{kl}(t) \geq 0.2$ for $l \in L$. Therefore, $0 \leq L_k \leq L$. The coverage criterion is satisfied if $L_k > \frac{L}{2}$. We define the key mutation set $W$ as follows:

$$W = \{ X_k \mid \exists l \; s.t., p_{kl}(t) \geq 0.5; \; L_k > \frac{L}{2} \}, \; k \in K, \; t \in T, \; l \in L. \qquad \textbf{Eq.1}$$

A total of 170 key mutations of SARS-CoV-2 variants were identified using these criteria and are listed in **Table S2** for further analysis.

### Genetic sequences

We retrieved 6.6 million complete SARS-CoV-2 genomes, isolated between March 2020 and June 2024, from the GISAID EpiCoV database[9] (accessible at gisaid.org/EPI_SET_250213es). Strains with low coverage duplicate names or ambiguous collection dates were filtered out. The remaining sequences were aligned using MAFFT (version 7)[17].

**The Equal Power Sampling (EPS) framework**

Our goal is to track the spread of mutations across space and time using serial cross-sectional data from multiple geographic locations. To prevent overrepresentation of countries with higher sequencing volumes as potential origins, we apply an equal power constraint to each data block for detecting low-frequency variants. Let $\rho$ denote the population mutation prevalence, and $N_\rho$ the sample size required to estimate $\rho$ with 90% power and 5% type I error rate. $N_\rho$ can be derived using established power estimation methods[18,19]. For instance, when $\rho = 0.005$, $N_{0.005} = 1,705$. Under the EPS framework, we set $N_\rho$ as the target sample size for each data stratum. As the prevalence of a key mutation increases following its emergence in a given region, a smaller sample size becomes sufficient in later time periods. Accordingly, the framework operates in two stages based on the level of $\rho$: $N_{0.005} = 1,705$ ensures adequate power to detect a rare variant when a key mutation first emerges, while $N_{0.05} = 163$ provides sufficient power to estimate mutation prevalence in subsequent time points. When identifying the source region, if the available sample size is smaller than $N_{0.005}$, all samples from the stratum are included. The EPS procedure is repeated B times to minimize random sampling variation, with each random sample indexed by $b$ ($1 \leq b \leq B$). The two-stage design and sampling approach ensures that the source region estimation remain unbiased by the sample size disparities.

**Inferring geographical transmission pathways using rank statistics**

We employed rank statistics to infer the relative positions of geographic regions along the viral transmission pathway, moving beyond a focus on the precise timing of mutation detection. For each key mutation $X_k$, its prevalence across all regions and time points can be estimated within a randomly sampled dataset $b$ under the EPS framework, denoted as $p_{kl}^b(t)$. Let $L_k^b$ represent the number of regions where mutation $X_k$ is observed, where $1 \leq L_k^b \leq L$. We define $l_{bk}^{(r)}$ as the geographical region that is the $r^{\text{th}}$ in the sequence to first detect mutation $X_k$ with continuous circulation (referred to as the "first detection" region). This is defined by $p_{kl}^b(t) > 0$ and $p_{kl}^b(t + \Delta t) > 0$ for $\Delta t = 0,1,2,\ldots,$ and $1 \leq r \leq L_k^b$. Consequently, $l_{bk}^{(1)}$ represents the source region for mutation $X_k$. Conversely, let $r_{kl}^b$ denote the rank of region $l$ in detecting $X_k$ among the $L_k^b$ regions. When $L_k^b < L$, the rank is normalized as $r_{kl}^{'b} = [r_{kl}^b \frac{L}{L_k^b}]$. This normalization enables the joint analysis of all mutations. The average ranks across $B$ random samples are then computed for further analysis using:

$$R_{kl} = \left[\frac{1}{B}\sum_{b=1}^{B} r_{kl}^{'b}\right].$$

Let $n_l(r)$ denote the frequency with which a region is observed at a specific rank $r$ across $K$ mutations. This can be expressed as:

$$n_l(r) = \sum_{k=1}^{K} I\{R_{kl} = r\},$$

where $I\{\cdot\}$ is the indicator function. Thus, the probability of a region appearing at a given rank $r$ among all regions can be estimated by:

$$\Pr\{R_l = r\} = n_l(r)/\sum_{l=1}^{L} n_l(r) \qquad \textbf{Eq. 2}$$

This formulation enables the calculation of rank probabilities from observed frequencies. Therefore, the most probable source region is identified as the location with the highest probability of occupying the first position in the viral transmission pathway. This relationship can be expressed as:

$$l^{(1)} = \arg\max_{1 \leq l \leq L} \Pr\{R_l = 1\}. \qquad \textbf{Eq. 3}$$

Similarly, the most likely sink region is determined by:

$$l^{(L)} = \arg\max_{1 \leq l \leq L} \Pr\{R_l = L\},$$

and the most likely location for any rank order $r$ (where $1 \leq r \leq L$) is given by:

$$l^{(r)} = \arg\max_{1 \leq l \leq L} \Pr\{R_l = r\}. \qquad \textbf{Eq. 4}$$

Using this Site-based mutation dynamic analysis within the EPS framework (S-EPS), we estimated the probability distributions of rank orders for geographical regions along the viral transmission pathway. Given our focus on identifying the source and sink of virus migration, and considering the sensitivity of intermediate ranks, we grouped the transmission stages between source and sink into three broad categories: Intersection 1 or early intersection ($R_l \in [2,3]$), Intersection 2 ($R_l \in [4,7]$), and Intersection 3 ($R_l \in [8,12]$) or late intersection ($R_l \in [4,12]$).

**Prediction lead time of the source region and prediction probability**

Information from the source region can be used to predict future predominant mutations elsewhere. To demonstrate this, we first calculate the lead time of the source (LTS) for mutation detection. In the $b^{\text{th}}$ random sample, let $t_{bk}^1(\theta_1)$ denote the time when a mutation is first detected in the source region with a prevalence $p_{k1}^b(t) > \theta_1$. Similarly, let $t_{bk}^l(\theta_2)$ denote the time when the mutation $X_k$ is detected in region $l$ with a prevalence $p_{kl}^b(t) > \theta_2$. The LTS for mutation $X_k$ relative to region $l$ in $b^{\text{th}}$ random sample is defined as:

$$LTS_{bk} = t_{bk}^l(\theta_2) - t_{bk}^1(\theta_1).$$

The overall LTS is then computed as the sample average across all $K$ mutations and $B$ random samples:

$$LTS = \frac{1}{B \cdot K} \sum_{b=1}^{B} \sum_{k=1}^{K} LTS_{bk}. \qquad \textbf{Eq. 5}$$

Here, $\theta_1$ and $\theta_2$ are prevalence thresholds within the range (0,1]. In our main results, $\theta_1 = 0.01$ serves a screening threshold for detecting a mutation in the source, while $\theta_2 = 0.50$ indicates that a mutation has reached predominance. Therefore, the probability that a mutation becomes predominant in region $l$ with a prevalence $\theta_2$, given its prevalence $\theta_1$ in the source region, is calculated as:

$$\Pr(p_{kl} > \theta_2 \mid p_{k1} > \theta_1) = \frac{\sum_{b=1}^{B} \sum_{k=1}^{K} I\{p_{kl} > \theta_2 \text{ AND } p_{k1} > \theta_1\}}{\sum_{b=1}^{B} \sum_{k=1}^{K} I\{p_{k1} > \theta_1\}}. \qquad \textbf{Eq. 6}$$

**Mutation transition time**

The mutation transition time ($\tau$) characterizes the duration required for a mutation's fitness to increase within a population until it first reaches a dominant prevalent threshold ($\theta$)[20], which in this study is set at $\theta = 0.50$. Let $t_{bkl}^0$ denote the time at which mutation $X_k$ is first observed to emerge and begin circulating continuously in the $b^{\text{th}}$ random sample from region $l$. This is defined as the moment when $p_{kl}^b(t) = 0$ and $p_{kl}^b(t + \Delta t) > 0$, with $\Delta t = 0,1,2,\dots$ Similarly, let $t_{bkl}^\theta$ denote the time when $p_{kl}^b(t)$ first reaches the threshold $\theta$, such that $p_{kl}^b(t_{bkl}^\theta + \Delta t) > \theta$. The transition time for mutation $X_k$ in the $b^{\text{th}}$ random sample in region $l$ is then calculated as:

$$\tau_{bk}^l = t_{bkl}^\theta - t_{bkl}^0.$$

The overall transition time for mutations in region $l$ is obtained by averaging the transition times across all random samples and mutations:

$$\tau^l = \frac{1}{B \cdot K} \sum_{b=1}^{B} \sum_{k=1}^{K} \tau_{bk}^l. \qquad \textbf{Eq. 7}$$

## Acknowledgments

We thank the GISAID Initiative for the surveillance efforts and open data sharing, and appreciate the data submitting laboratories. This work was supported by the National Natural Science Foundation of China: 32322088 (M.H.W.), Research Grants Council of Hong Kong (Theme-Based Research Scheme: T11-705/21-N; L.L.M.P.), the Chinese University of Hong Kong (CUHK) Research Committee - Strategic Seed Funding for Collaborative Research Scheme 2022-23 (M.H.W.), and CUHK Direct Grant 2022.02 (M.H.W.).

## Author Contributions Statement

M.H.W. conceived the study, acquired funding and supervised the project. M.H.W. and H. Z. designed the method and wrote the manuscript. H.Z. wrote code and analyzed output data. S.S., J.L., C. L. L.C. and Y.Z contributed to data collection and interpretation. M.K.C.C., B.C.Y.Z., P.P.H.C., H.G., J.P., L.L.M.P. and H.L.Y. revised the manuscript.

## Competing Interests Statement

The Chinese University of Hong Kong filed a pending patent (US Provisional application no. 63/632,442) covering the method described in this paper, listing M.H.W., S.S., H.Z., J.L., and B.C.Y.Z as the inventors. M.H.W. and B.C.Y.Z. are shareholders of Beth Bioinformatics Co., Ltd. B.C.Y.Z. is a shareholder of Health View Bioanalytics Ltd. The remaining authors declare no competing interests.

**Figures**

**Figure 1. Global circulation pathway of SARS-CoV-2 variants**

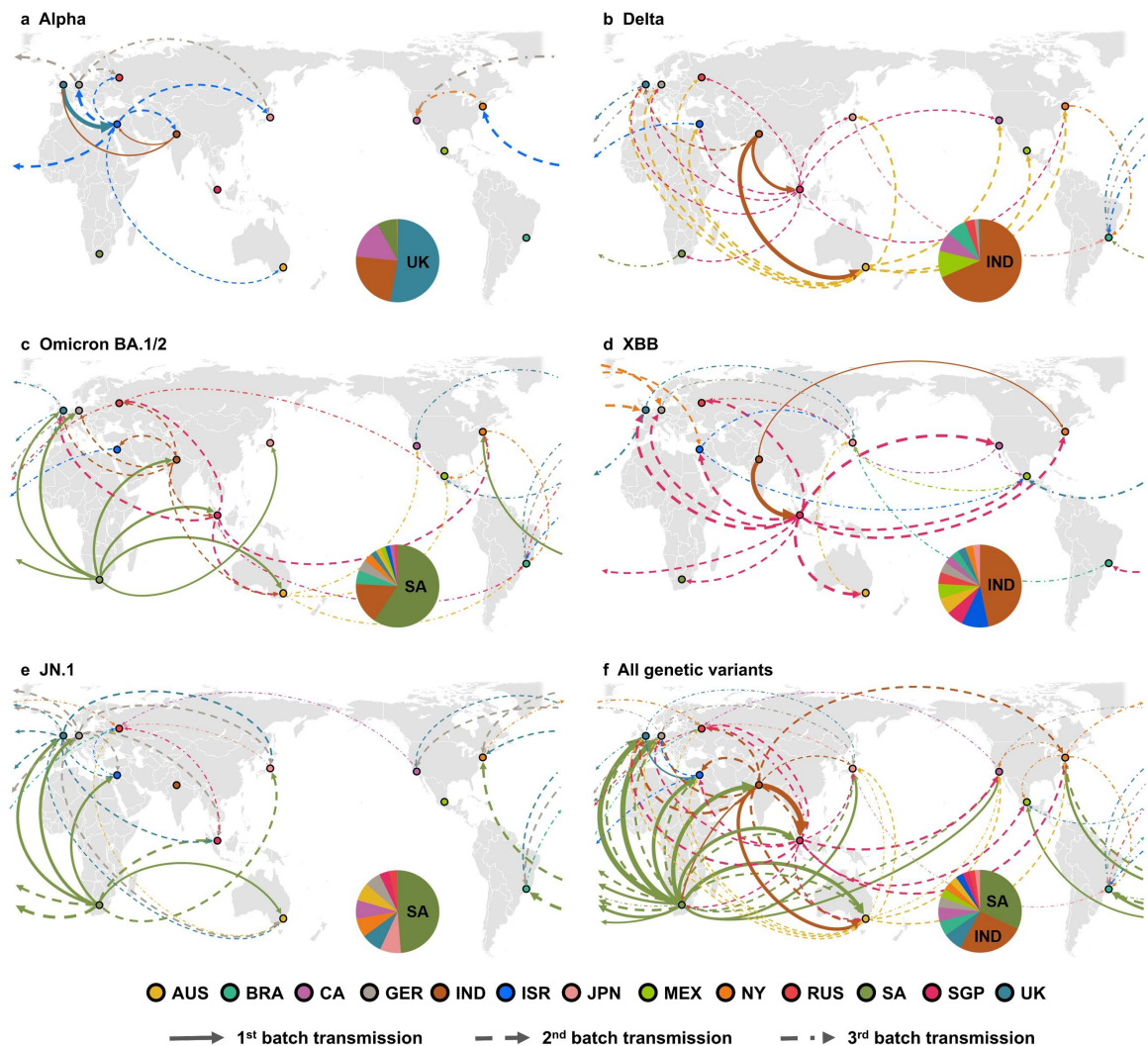

**Legend:** Global circulation pathway of SARS-CoV-2 variants inferred by the probability of geographical regions' rank statistics using key mutations of **a** Alpha. **b** Delta. **c** Omicron BA.1/2. **d** Omicron XBB and its sublineages. **e** Omicron JN.1 and its sublineages. **f** All genetic variants. The probabilities of the 13 geographical regions being the source of a genetic variant are plotted in the pie chart. 1st batch transmission: from the source region (rank No. 1) to early intersection regions (rank No. 2-3), 2nd batch transmission: from early intersection to late intersection regions (rank No. 4-12), 3rd batch: from late intersection regions to sink (ranked No. 13). Width of the transmission line is proportional to the number of mutations following the routes. AUS: Australia, BRA: Brazil, CA: California, GER: Germany, IND: India, ISR: Israel, JPN: Japan, MEX: Mexico, NY: New York, RUA: Russia, SA: South Africa, SGP: Singapore, UK: United Kingdom.

**Figure 2. Time and location of initial detection of the SARS-CoV-2 key mutations**

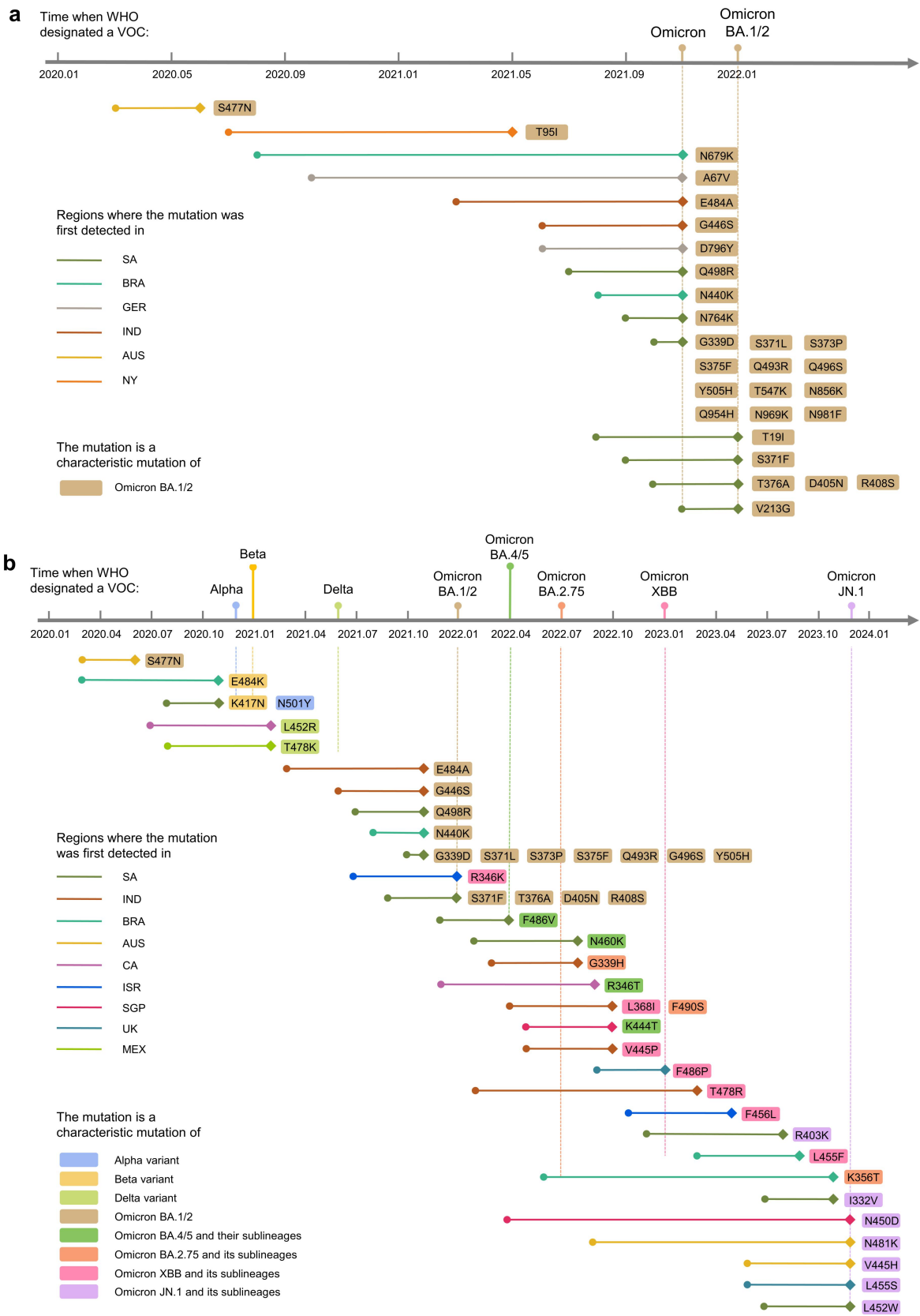

**Legend:** Time and location of initial detection of the key mutations of **a** S protein of Omicron BA.1/2; **b** Receptor binding domain (RBD) of all major genetic variants of SARS-CoV-2. Diamond ♦: the mutation reaches predominance in any one of the 13 geographical regions. This figure shows that the key mutations of each SARS-CoV-2 variants emerged in multiple geographical regions, while the country reporting the initial epidemic outbreak of a variant played a critical role by contributing the last and largest batch of key mutations of the variant.

**Figure 3. Probability of geographical regions' position along global circulation pathway of SARS-CoV-2**

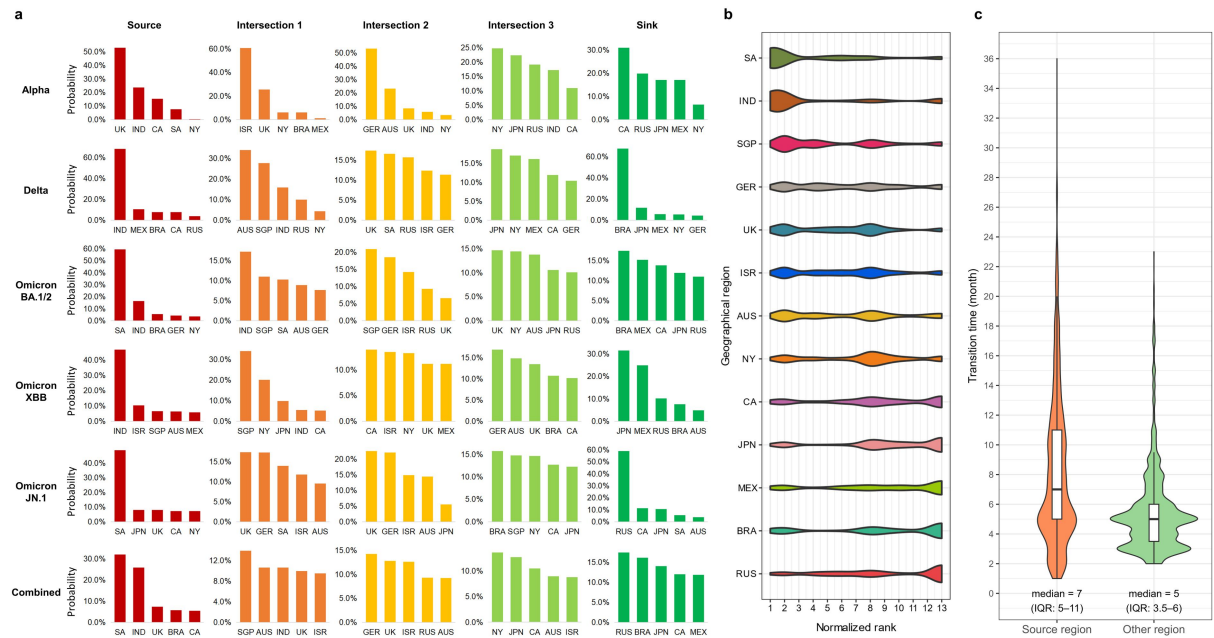

**Legend**: **a** Geographical regions' probability of being at a given position (source, intersections, sink) along the SARS-CoV-2 global circulation pathway. Source (region ranking No. 1): the first place of observing key mutations out of the 13 geographical regions studied. Intersection 1 or early intersection (regions ranking No. 2-3), Intersection 2 (ranking No. 4-7), Intersection 3 (ranking No. 8-12), and the Sink (ranking No. 13). For a given position, the top five regions with the highest probability are shown. The 13 geographical regions investigated include, AUS: Australia, BRA: Brazil, CA: California, GER: Germany, IND: India, ISR: Israel, JPN: Japan, MEX: Mexico, NY: New York, RUA: Russia, SA: South Africa, SGP: Singapore, UK: United Kingdom. Over all variants investigated between March 2020 and June 2024, the most probable genetic source of SARS-CoV-2's key mutations are estimated to be Africa (South Africa) and India subcontinent (India). **b** Distribution of ranks of geographical regions along viral transmission pathway. **c** Distribution of mutation transition time in the source and outside the source.